\documentclass[aps,10pt,prb,
groupedaddress]{revtex4}

\usepackage{graphicx}

\usepackage[caption=false]{subfig}
\usepackage{amsmath}
\usepackage{bbold}
\usepackage{latexsym}

\begin{document}


\title{Why a noninteracting model works for shot noise in fractional charge experiments}


\author{D. E. Feldman$^1$ and Moty Heiblum$^2$}
\affiliation{$^1$Department of Physics, Brown University, Providence, Rhode Island 02912, USA}
\affiliation{$^2$Braun Center for Submicron Research, Department of Condensed Matter
	Physics, Weizmann Institute of Science, Rehovot, 76100, Israel}


\date{\today}

\begin{abstract}
A fractional quasiparticle charge is a manifestation of strong interactions in the fractional quantum Hall effect. Nevertheless, shot noise of quasiparticles is well described by a formula, derived for noninteracting charges. We explain the success of that formula by proving that in the limits of strong and weak backscattering it holds irrespectively of microscopic details in weakly and strongly interacting systems alike. The derivation relies only on principles of statistical mechanics. We also derive an approximate model-independent formula for shot noise in the regime of intermediate backscattering. The equation is numerically close to the standard `noninteracting' fitting formula but suggests a different physical interpretation of the experimental results. We verify our theoretical predictions with  a shot noise experiment at the filling factor $3/5$.
\end{abstract}



\maketitle


\section{Introduction}

One of the most interesting features of topological matter is fractionalization.  In non-topological materials we usually expect that the quantum numbers of excitations are multiples of the quantum numbers of constituent particles.
For example, elementary excitations of Fermi liquids carry the charge and spin of an electron. A composite of $n$ electrons has the charge $ne$ and its spin is the sum of $n$ spins $1/2$. Fractionalization allows the excitation quantum numbers to deviate from the combinations of the quantum numbers of constituent particles. In many cases, the former are fractions of the latter. In others, the former violate constraints, which must be satisfied by any finite groups of constituent particles.  A famous example is the spin-charge separation in the Su-Schrieffer-Heeger model of polyacetylen \cite{ssc}, where domain wall solitons can carry an electron charge $e$ with zero spin or zero charge with spin $1/2$. Spin-charge separation was also predicted in topological insulators \cite{s-c} and in non-Abelian quantum Hall states \cite{mr,apf1,apf2,php}, where an electron can be seen as a combination of a charged boson and a neutral Majorana fermion \cite{sss}. Another type of fractionalization is encountered in spin ice, where spins fractionalize into magnetic monopoles \cite{s-f}. A review of many fractionalization phenomena can be found in Ref. \onlinecite{ds}. The simplest and most striking example is a fractional electric charge, observed in the quantum Hall effect \cite{tQHE}. 

A fractional charge was an early key prediction in the theory of the fractional quantum Hall effect \cite{Laughlin}.  The prediction of charge-$e/3$ quasiparticles at the filling factor $1/3$ was confirmed spectacularly in the 90s, shot noise providing the most useful tool  \cite{fc1,glattli}. Subsequent experiments brought a wealth of information about fractional charges at odd- and even-denominator quantum Hall plateaus
 \cite{fc1,glattli,noise1,noise2,noise3,noise-5/2,5/2-1,5/2-2,Bid,dolev,charge-int,noise4,heiblum}. In particular, the observation \cite{noise-5/2,5/2-1,5/2-2} of the charge $e/4$ at $\nu=5/2$ was an important step towards understanding the nature of a potentially non-Abelian $5/2$ state. 

Shot noise experiments have stimulated much theoretical work \cite{KF,FLS1,FLS2,FLS3,FS,WF1,WF2,WF3,LS,Tr,T1,Ferraro,5-2,T2,T3,T4,T5,T6,T7,ups,CT1,CT2,CT3,CT4}. A basic theoretical framework to understand the experiments is based on the chiral Luttinger liquid model with a point scatterer \cite{KF}. The model admits an exact solution that shows clear signatures of fractionally charged excitations \cite{FLS1,FLS2,FLS3,FS}. However, that solution turns out to provide a poor fit to the current and noise data \cite{heiblum}. Instead, a simple formula, derived for noninteracting fermions, is routinely used to fit the data \cite{heiblum}. The success of that model is puzzling given strongly interacting nature of the fractional quantum Hall physics. The goal of this paper is to shed light on that success.

\begin{figure}[b]
	\centering
	\includegraphics[width=4in]{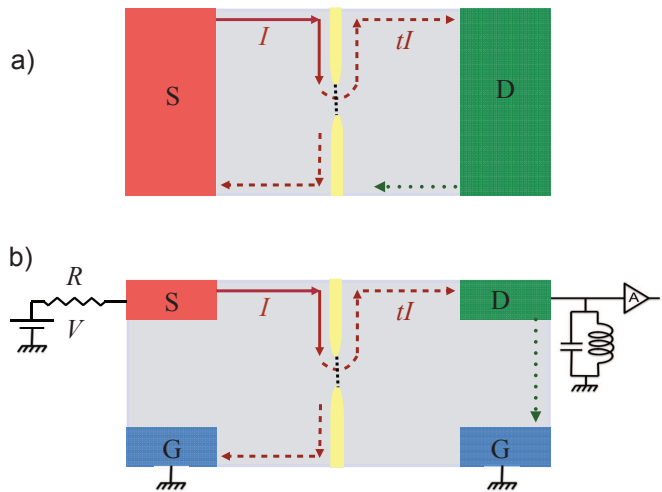}
	\caption{(color online) Two setups for the measurement of the fractional charge. Current flows along the edges in the direction of the arrows. It tunnels between the edges along the dotted line at the quantum point contact. The amplifier measures the voltage noise with respect to the ground in drain D.}
	\label{Fig1}
\end{figure}

Figure 1 shows two setups used in charge measurements in the quantum Hall regime. We will focus on the four-terminal setup b), which is typically used in the experiments, while the papers on the chiral Luttinger liquid model \cite{KF,FLS1,FLS2,FLS3,FS} all focused on setup a). The equations for the noise are similar but not the same in the two geometries. This is because the currents leaving and entering the drain along the chiral channels in Fig. 1a) both have non-zero cross-correlations with the tunneling current at the QPC. This is not the case in setup b), where the edge channel leaving drain D terminates in a grounded ohmic contact. Only minor modifications are needed to connect our results with geometry a).

In a most general situation, $n$ integer channels are fully transmitted by the QPC and one more integer or fractional channel is partially transmitted. We will focus on $n=0$ below. The extension of our results to $n>0$ is straightforward.

Three quantities are measured experimentally. The first is the average transparency $t$ of the QPC. It is defined as $t=I_D/I_S$, where  $I_D$ is the drain current, transmitted by the QPC, $I_S=GV$ is the impinging source current at the bias voltage $V$ and the quantized Hall conductance $G=\nu e^2/h$. The second measured quantity is the differential conductance $t_d=dI_D/d I_S=d I_D/GdV$. In many cases it is close to the average transparency
$t$. We will see that at $t\ne t_d$ the noise depends on both transparencies.
The third measured quantity is the spectral density of the shot noise in the drain at zero frequency,

\begin{equation}
\label{noise-1}
S=\int_{-\infty}^\infty dt[\langle \hat I_D(0)\hat I_D(t)+\hat I_D(t) \hat I_D(0)\rangle- 2\langle \hat I_D\rangle^2].
\end{equation}
The fractional charge ${\tilde e}$, partitioned at the QPC, is extracted from the equation \cite{heiblum}

\begin{equation}
\label{noise-2}
S=2{\tilde e} V G t(1-t)\left[\coth\left(\frac{{\tilde e}V}{2k_BT}\right)-\frac{{2k_BT}}{{\tilde e} V}\right]+4k_BTG.
\end{equation}
An equation of such type with ${\tilde e}=e$ was derived for noninteracting electrons in Ref. \onlinecite{ML}. However, the fractional QHE is an example of an interaction-driven phenomenon. Thus, the success of Eq. (\ref{noise-2}) in the interpretation of experiments is a puzzle.

We will solve that puzzle in two steps. First, we consider the regimes of small and large transparency: $t\approx 0$ and $t\approx 1$. In both cases, we establish Eq. (\ref{noise-2}) 
solely from the basic principles of statistical mechanics without any reference to a particular model. We next consider the opposite limit of $t=1/2$. In that case no exact model-independent formula exists but we discover that after a small modification, Eq. (\ref{noise-2}) holds with a relative error of $\sim[t(1-t)]^2$. This explains why Eq. (\ref{noise-2}) fits the data even at $t=1/2$ with $\tilde e$, intermediate between the electron charge and the fractional quasiparticle charge. A shot noise experiment at the filling factor $\nu=3/5$ verifies the theory.

\section{Small and large $t$}.

Our starting point is one of the fluctuation-dissipation relations, proven in Refs. \onlinecite{WF1,WF2,WF3}. The relation holds for any chiral system, such as a QHE edge, and its derivation does not depend on a model. Note that the relevant fluctuation-dissipation theorem has long been known to hold for the exact solution of the chiral Luttinger liquid model \cite{KF,FS}. In geometry 1b) the theorem states 

\begin{equation}
\label{noise-3}
S=S_Q-4k_BT\frac{\partial I_Q^q}{\partial V}+4k_BTG,
\end{equation}
where $I_Q^q=(1-t)G V$ is the current, reflected at QPC, and its noise $S_Q=\int_{-\infty}^\infty dt[\langle \hat I_Q^q(0)\hat I_Q^q(t)+\hat I_Q^q(t) \hat I_Q^q(0)\rangle- 2\langle \hat I_Q^q\rangle^2]$.

Let us start with the limit of small $(1-t)$ and thus small $I_Q^q$.
The third term of Eq. (\ref{noise-3}) is the same as in Eq. (\ref{noise-2}). The first term of Eq. (\ref{noise-3}) is also the same as the first term $2{\tilde e} V G  t(1-t)\coth\left(\frac{{\tilde e}V}{2k_BT}\right)$ in the brackets in Eq. (\ref{noise-2}) with ${\tilde e}$ set to be the quasiparticle charge $e^*$. This follows from the detailed balance principle, as shown in Ref. \onlinecite{LS}.
We rely on the standard assumption that rare tunneling events of $e^*$ quasiparticles at QPC are uncorrelated \cite{LS}. Let $p_1$ be the rate of such tunneling events from the upper to lower edges in Fig. 1. The rate $p_2$ of the reverse tunneling process from the lower to upper edge is given by the detailed balance relation: $p_2=p_1\exp(-e^*V/k_B T)$. The tunneling current $I_Q^q=e^*(p_1-p_2)$. To find the noise $S_Q$ we observe that $S_Q=2\langle \Delta Q^2 \rangle/t_{\rm long}$, where $\Delta Q$ is the fluctuation of the charge, backscattered during a long time $t_{\rm long}$.
We express $\Delta Q$ as $\Delta Q=\sum_i \Delta Q_i$, where $\Delta Q_i$ are uncorrelated fluctuations of the backscattered charge during short time intervals $\Delta t_i$.
Thus, 

\begin{eqnarray}
\label{noise-8}
\langle \Delta Q^2 \rangle = \sum_i \Delta Q_i^2= & & \nonumber\\
\sum_i \left[ (p_1+p_2)(e^*)^2 \Delta t_i-(p_1-p_2)^2(e^*)^2\Delta t_i^2\right] & & \nonumber\\
\approx (p_1+p_2)(e^*)^2t_{\rm long}. 
\end{eqnarray}
A comparison with the expression for the current
shows that $S_Q=2e^* I_Q^q (p_1+p_2)/(p_1-p_2)=2e^* I_Q^q \coth \frac{e^*V}{2k_B T}$ in agreement with Eq. (\ref{noise-2}) and Ref. \onlinecite{LS}.
Thus, in the limit of large transmission $t\approx 1$, Eq. (\ref{noise-3}) becomes

\begin{equation}
\label{noise-4}
S=2e^*VG (1-t)\coth\frac{e^* V}{2k_B T}-4k_B T\frac{\partial I_Q^q}{\partial V}+4k_B TG.
\end{equation}

The above expression simplifies if the transmission $t$ does not depend on the bias voltage $V$. Under such an assumption,
the second term in Eq. (\ref{noise-4}) becomes $4k_B T (1-t)G  $ in agreement with the second term in the brackets in Eq. (\ref{noise-2}) at $t\approx 1$.
Thus, we have derived Eq. (\ref{noise-2}) for the fractional charge  from the principles of statistical mechanics at weak backscattering. The only assumption was the voltage independence of $t$. 
Note that the above assumption may not describe the experimental data well \cite{heiblum}. In such a case, the average transmission $t$ and and the differential transmission $t_d$ differ and it has been known that the agreement with the fitting formula (\ref{noise-2}) could be improved by using a combination of  $t$ and $t_d=1-{\partial I_Q^q}/G {\partial V}$ in place of $t$. We can understand that from Eq. (\ref{noise-4}). Indeed, Eq. (\ref{noise-4}) can be rewritten as

\begin{equation}
\label{noise-2-add}
S_D=2e^*VG  t(1-t)\coth\frac{e^* V}{2k_B T}-4k_B T G  t_d(1-t_d)+4k_B TG,
\end{equation}
where both the average transparency $t\approx 1$ and the differential transparency $t_d \approx 1$ enter and we rely on $t(1-t)\approx (1-t)$.  
Note that the equivalent Eq. (\ref{noise-4}) was found to fit the noise data in Ref. \onlinecite{glattli}. Equation (\ref{noise-2-add}) is a general model-independent result that applies to any QPC at any filling factor, voltage and temperature as long as $t\approx 1$.

So far we relied on a physical picture of quasiparticle tunneling in the up-down direction in the QPC in Fig. 1b). To address the limit of small $t \ll 1$ we change
our interpretation of Fig. 1b) and think in terms of electron tunneling. In such a physical picture, 
electrons and not fractional quasiparticles tunnel at QPC  in the right-left direction with a small backscattering rate $t$. This change of the qualitative picture reflects the fact that at small $t$ the QPC separates two quantum Hall liquids with a non-topological insulator in between. Fractional quasiparticles are excitations of a strongly-correlated electron system. They cannot exist without an underlying topological liquid. In particular, the only charged excitations of a non-topological insulator are electrons. Hence, charge tunneling through the QPC can only be mediated by electrons.

The same arguments as at $t\approx 1$ yield Eq. (\ref{noise-2}) with the effective charge ${\tilde e}=e$. Again, the only assumption is the voltage independence of $t$. If that assumption does not hold then a modified Eq. (\ref{noise-4}) should be used:

\begin{equation}
\label{noise-5}
S_D=2eVG  t\coth\frac{e V}{2k_B T}-4k_B T\frac{\partial I_Q^e}{\partial V}+4k_B TG,
\end{equation}
where $I_Q^e=I_D=tG V$ is the tunneling current of electrons and $t=t(V)$ is the average transparency. The direction of the tunneling current is from the left to the right in Fig. 1b).  Equation (\ref{noise-5}) can be rewritten in the form, similar to (\ref{noise-2-add}), with both the average and differential transparencies.

The above derivation relies on the fluctuation-dissipation theorem \cite{WF1,WF2,WF3} whose proof assumes chiral edges, such as the edges at the filling factors $1/3$ and $2/5$ but not $2/3$ and $3/5$. For a judiciously chosen experimental setup, our results also hold for nonchiral edges as discussed in the Appendix.

\section{Transparency $t=1/2$}

The previous section derives Eq. (\ref{noise-2}) in the leading order in the small $t$ or small $(1-t)$.
One can thus expect the relative accuracy of the  tunneling charge, determined from Eq. (\ref{noise-2}) in the limit of small $t$, to be of the order of $t$. In the limit of small $1-t$ the accuracy is $\sim (1-t)$. The experimental accuracy is generally worse than 10\% and thus the above analysis
justifies the use of Eq. (\ref{noise-2}) at $t<0.2$ and $t>0.8.$ 
Yet, Eq. (\ref{noise-2}) was used with success for all values of $t$. A comparison with the exact solution for the chiral Luttinger liquid model \cite{Tr} also showed an agreement with Eq. (\ref{noise-2}) in a broad range of $t$. 

To understand the surprising power of Eq. (\ref{noise-2}) we focus on the limit of $t=1/2$,   which corresponds to the maximal noise. Indeed, $t=1$ means a fully transparent QPC with no quasiparticle tunneling in Fig. 1b) and zero noise. At $t=0$ there is no electron tunneling in the dual picture of the tunneling current from the left to the right. The noise $S$ depends on the voltage $V$, temperature $T$ and the split-gate voltage $V_G$, which controls the transmission $t=t(V_G,V,T)$. We can thus treat $S$ as a function of $V,T$ and $t$: $S=S(V,T,t)$. In what follows we will assume that $t$ does not depend strongly on $V$ at fixed $T$ and $V_G$, and hence, Eq. (\ref{noise-2}) can be used at small $t$ and small $1-t$. In the former limit one substitutes $e$ for ${\tilde e}$. In the latter limit $\tilde e=e^*$.

Let us consider the combination ${\tilde {S}}=[S(t,V,T)+S(1-t,V,T)]/2$.
 In the case of interest, $t=1/2$, the combination $\tilde {S}$ reduces to $S$. Since ${\tilde S}(t)={\tilde S}(1-t)$, we observe that the power expansion of $\tilde S$ about $t=1/2$ contains only even powers of $(t-1/2)$. In other words, $\tilde S$ can be expanded in integer powers of $t(1-t)=1/4-(t-1/2)^2$, $\tilde S=S_0+t(1-t)S_1+[t(1-t)]^2S_2+\dots$. 
Neglecting $S_2$ and the higher order terms brings a relative error $\sim t(1-t)$ in comparison with the $S_1$ contribution. Even at $t=1/2$ the relative error $t(1-t)$ is estimated as 25\% while the accuracy of the experiments does not generally exceed 10\%. We thus drop all terms beyond $S_1$. 

We still need to find $S_0$ and $S_1$. This can be done by looking at small $t$ [or, equivalently, large $t\approx 1$], where Eq. (\ref{noise-2}) holds. We thus discover that

\begin{eqnarray}
\label{noise-6}
{\tilde S}=G  t(1-t)\left[ e^*V\coth\left(\frac{e^*V}{2k_BT}\right)
+eV\coth\left(\frac{eV}{2k_BT}\right)\right] & & \nonumber \\
-4G t(1-t)k_B T
+4k_BTG,
\end{eqnarray}
where $e$ is an electron charge and $e^*$ is a quasiparticle charge. The tunneling operator, most relevant in the renormalization group sense at $t\approx 1$, may transfer the charge $ne^*$ of $n>1$ quasiparticles. At the same time, the electrostatic barrier increases with the charge of the tunneling object. The dominant tunneling process at given $V, T$ and $t$ is determined by the unrenormalized high-energy transparency and its low-energy renormalization at the scale ${\rm max}(e^*V,k_BT)$. The unrenormalized transparency is maximal for charge-$e^*$ particles. We thus expect that in the limit $t\rightarrow 1$ at fixed $V$ and $T$ the dominant tunneling process involves one lowest-charge quasiparticle, such that $e^*=e/3$ at $\nu=1/3$ and $e^*=e/5$ at $\nu=3/5$.

\begin{figure}[b]
	\centering
	\includegraphics[width=8in]{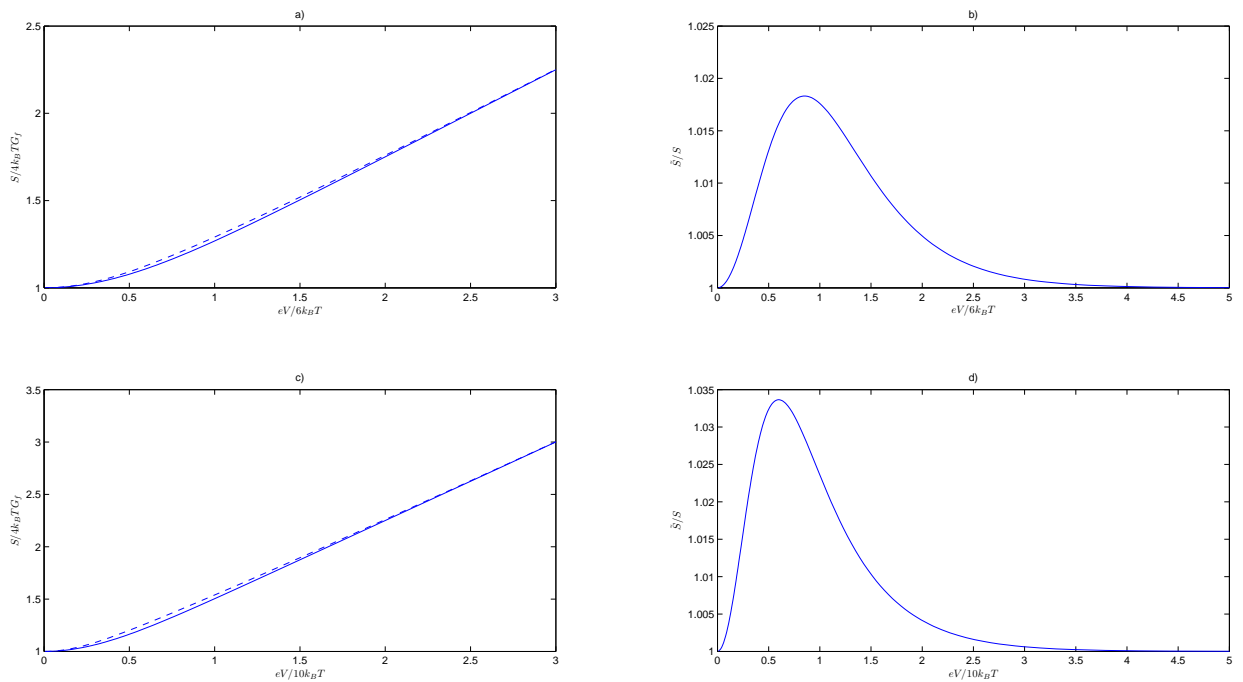}
	\caption{(color online) a) The solid line shows the standard fitting formula (\ref{noise-2}) at $\nu=2/3$ with $t=1/2$, $G =2e^2/3h$ and $\tilde e=2e/3$. The dashed line is (\ref{noise-6}) with $e^*=e/3$.
b) The ratio of (\ref{noise-6}) and (\ref{noise-2}) at the same parameters as in a). c) The solid line is (\ref{noise-2}) at $\nu=3/5$ with $t=1/2$, $G =3e^2/5h$ and $\tilde e=3e/5$. The dashed line is (\ref{noise-6}) with $e^*=e/5$. d) The ratio of (\ref{noise-6}) and (\ref{noise-2}) at the same parameters as in c).}
	\label{Fig_plots}
\end{figure} 

At $t=1/2$ the above expression (\ref{noise-6}) describes the measured noise $S$. While Eq. (\ref{noise-6}) is not exactly the same as Eq. (\ref{noise-2}), it is very similar. The only difference of the two equations consists in the number of $\coth$ terms.  Moreover, the expression in the square brackets in (\ref{noise-6}) does not differ much from the $\coth$ contribution to Eq. (\ref{noise-2}) if an appropriate effective charge $\tilde e$ is substituted into Eq. (\ref{noise-2}). Indeed, at large $V$, the voltage-dependent contribution to Eq. (\ref{noise-2}) reduces to $2{\tilde e}VG  t(1-t)$. The voltage-dependent part of Eq. (\ref{noise-6}) assumes exactly the same form for ${\tilde e}=
(e+e^*)/2$. At low voltages, the voltage dependence of Eq. (\ref{noise-2}) becomes
$\frac{G t(1-t)}{3k_BT} ({\tilde e}V)^2$. From Eq. (\ref{noise-6}) we get
$\frac{G t(1-t)}{6k_BT}[(eV)^2+(e^*V)^2]$ at low voltages. These two expressions become the same at ${\tilde e}=\sqrt{[e^2+(e^*)^2]/2}$. At first sight there is a problem: the expressions for $\tilde e$ are not identical in the high and low voltage limits. However, the difference does not exceed the experimental accuracy. For example, at $e^*=e/3$, the two estimates for $\tilde e$ are $2e/3$ and $\sqrt{5}e/3$. Their difference is just 10\%. We emphasize a small numerical difference of equations (\ref{noise-2}) and (\ref{noise-6}) at all voltages and temperatures as illustrated by Fig. 2 for $e^*=e/3$ and $e^*=e/5$.

\section{Comparison with the data}

We now compare Eq. (\ref{noise-6}) with the noise data at the filling factor $\nu=3/5$. The choice of the filling factor is due to the existence of a robust plateau in the transmission $t$ at $t=t_p=5/9$.
That value differs from $t=t_0=1/2$ assumed in Eq. (\ref{noise-6}), yet the difference of $t_p(1-t_p)=20/81$ and $t_0(1-t_0)=1/4$ is just $1\%$ and can be neglected. Since the transmission does not depend on the bias voltage, the distinction between the average and differential transmission becomes irrelevant.

\begin{figure}[b]
	\centering
	\includegraphics[width=4in]{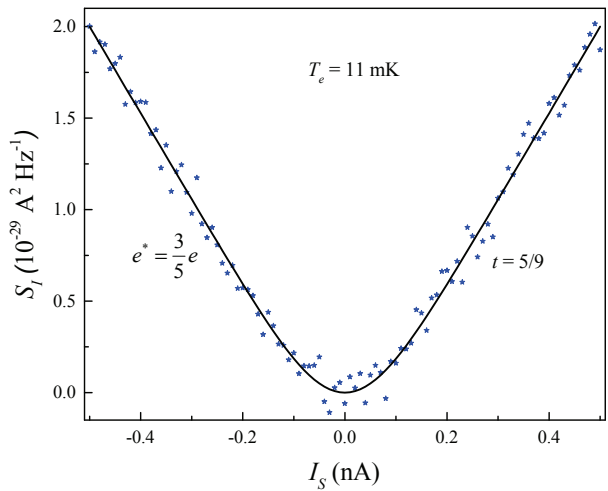}
	\caption{(color online) The excess shot noise at $\nu=3/5$ and $t=5/9$ vs the source current $I_S=3e^2V/5h$.}
	\label{fig_4_noise}
\end{figure}

The experiment was performed with a GaAs-AlGaAs heterostructure, embedding a 2D electron gas 130 nm below the surface. The areal density of the gas was $0.88\times 10^{11}~{\rm cm}^{-2}$
and the mobility $4.6\times 10^6~{\rm cm}^2{\rm V}^{-1}{\rm s}^{-1}$. 
 Annealed Au/Ge/Ni was used for the ohmic contacts.
The QPC was defined by split-gates with the opening of 850 nm. 
The gates were made with electron beam lithography followed by the deposition of Ti/Au.  The electron temperature was verified to be 11 mK. 

The noise signals  were filtered with an LC circuit
tuned to 700 kHz. The signals were first amplified by a cooled home-made preamplifier with voltage
gain 5. Its output was then sent to a room temperature amplifier  NF-220F5 with voltage gain 200, followed by a spectrum analyzer with the bandwidth of  20 kHz.

The voltage dependence of the noise is shown in Fig. 3. We used the standard fitting procedure with Eq. (\ref{noise-2}). The extracted tunneling charge $\tilde e=0.6e$. This compares favorably with the prediction (\ref{noise-6}). Indeed, the high voltage behavior of Eq. (\ref{noise-6}) is the same as that of Eq. (\ref{noise-2}) for the fitting charge
$\tilde e=(e+e/5)/2=0.6e$. The low voltage asymptotic agrees best with Eq. (\ref{noise-2}) at $\tilde e=\sqrt{(e^2+[e/5]^2)/2}=0.7e$, which cannot be meaningfully distinguished from 
$0.6e$ within the accuracy of Eq. (\ref{noise-6}).

Our derivation of Eq. (\ref{noise-6}) is based on the assumption that $t\approx t_d$ at small and large $t$: $t\ll 1$ and $(1-t)\ll 1$. This assumption is not valid in some intervals of voltages and temperatures. Yet, the agreement of our results and the standard fitting formula, Fig. 2, is excellent. To understand that we focus on two limitting cases: $eV\ll k_B T$ and $eV\gg k_B T$. Transport is linear in the former case and hence our assumption holds. In the latter case, the dominant contribution to the noise at large  $t$ comes from the first term in Eq. (\ref{noise-2-add}) and the dominant contribution to the noise at small $t$ is given by the first term in Eq. (\ref{noise-5}). Hence, substituting $t$ for $t_d$ in Eqs. (\ref{noise-2-add},\ref{noise-5}) does not create significant mistakes even though $t_d$ is not necessarily close to $t$. This justifies our approach.
 
\section{Conclusion}

The excellent agreement of Eq. (\ref{noise-6}) with the data should not be taken too seriously. First of all, Eq, (\ref{noise-6}) is a rather crude approximation with the
expected accuracy $\sim 25\%$. Besides, the experiment was performed at $t=5/9$, which is above $t=1/2$ assumed in the theoretical analysis. Yet, Eq. (\ref{noise-6}) has no fitting parameters and its success shows that the equation captures essential physics. It also sheds light on the success of Eq. (\ref{noise-2}), which is close to Eq. (\ref{noise-6}) numerically. Most importantly, Eq. (\ref{noise-6}) suggests a different interpretation of the data. The most natural interpretation of the charge ${\tilde e}$, extracted from fitting the data with Eq. (\ref{noise-2}), is a picture of tunneling objects with charge ${\tilde e}$. Unless ${\tilde e}$ is an integer multiple of the charge of a quasiparticle or electron the nature of such objects is puzzling. Eq. (\ref{noise-6}) suggests the interpretation of the same data in terms of a competition between electron and quasiparticle tunneling. At small $t$ electron tunneling wins and at large $t\approx 1$ quasiparticle tunneling wins. Such an intuition leads to a new formula for fitting the data. We propose an equation that interpolates between Eq. (\ref{noise-2}) at $t\ll 1$, $(1-t)\ll 1$ and Eq. (\ref{noise-6}) in the intermediate regime:

\begin{eqnarray}
\label{noise-7}
S=2G  t(1-t)\left[ pe^*V\coth\left(\frac{e^*V}{2k_BT}\right)
+(1-p)eV\coth\left(\frac{eV}{2k_BT}\right)\right] & & \nonumber \\
-4G t(1-t)k_B T
+4k_BT G,
\end{eqnarray} 
where the fitting parameter $p$ determines the dominant tunneling process. At $p=1$ that process is quasiparticle tunneling and at $p=0$ that is the tunneling of electrons. It may happen that the most relevant tunneling process at high transparency $t\approx 1$ involves quasiparticle bunching. For example, it was argued that the dominant tunneling process involves quasiparticle pairs \cite{Bid,Ferraro,5-2} at $\nu=2/3$. In such a case, a quasiparticle pair charge should be used in place of $e^*$ in the above formula. After such a modification, the experimental results \cite{Bid} at $t=1/2$ agree with our theory within its accuracy at $\nu=2/3$. In a most general case, several different tunneling charges compete and Eq. (\ref{noise-7}) should include contributions from them all. The importance of the competition between different quasiparticle types for tunneling in quantum Hall devices was emphasized in Refs. \onlinecite{T1,Ferraro,5-2,T2,T3,T4,T5}.

One generally expects that the tunneling of fractionally charged quasiparticles dominates at $t\approx 1$ and the electron tunneling dominates at $t\approx 0$. The experimental results of Ref. \onlinecite{dolev} disagree with such a picture at $\nu=7/3$ and $\nu=5/2$. Figure 4 of Ref. \onlinecite{dolev} also shows a surprising crossover between the electron tunneling at $t\approx 1$ and the quasiparticle tunneling at lower $t$ for $\nu=1/3$. Quasiparticle bunching may be responsible. Our equation (\ref{noise-2-add}) sheds additional light on the origin of such apparent behavior. Indeed, the tunneling charge $e$ appears in the regime, where the transmission is nonlinear and hence the standard fitting equation (\ref{noise-2}) does not work. Formula (\ref{noise-2-add}) should be used instead. Such an argument does not apply at $\nu=5/2$, where a large tunneling charge $0.8e$ is reported at a voltage-independent transmission $t\approx 1$ in Fig. 2a of Ref. \onlinecite{dolev}.  A similar behavior was observed in Ref. \onlinecite{noise4}, where charge $e/3$ was seen in the shot noise at intermediate transparencies at an integer filling factor and charge $e$ was seen at high and low transparencies. The physics behind the results \cite{noise4} differes from quasiparticle bunching. The point is that the quantum Hall liquid inside a device may  not be uniform. In particular, a $\nu=1/3$ region forms inside the QPC in an interval of gate voltages in the experiment \cite{noise4}. Only in that interval does the quasipartilce tunneling occur.

In conclusion, we used basic principles of statistical mechanics to derive a model-independent expression for the noise at low and high transmission of a QPC. For the voltage-independent transmission $t$ our results agree with the conventional fitting formula (\ref{noise-2}). A modified equation holds in the case of a voltage-dependent transparency. The analysis of a half-open QPC at $t=1/2$ sheds light on the success of Eq. (\ref{noise-2}) in the intermediate range of transmission $0<t<1$. The fractional charge, extracted from Eq. (\ref{noise-2}) in that range, does not describe the charge of the tunneling objects. This motivates an alternative fitting formula (\ref{noise-7}).

\acknowledgements

We thank M. Banerjee for very valuable discussions and providing the noise data. This research was supported in part by the
European Research Council under the European Community's Seventh Framework 
Program (FP7/2007-2013)/ERC Grant agreement No. 339070, the
Minerva foundation, grant No. 711752, and the German Israeli
Foundation (GIF), grant No. I-1241-303.10/2014.
DEF acknowledges the hospitality of the Weizmann Institute of Science.

\appendix

\section{The role of upstream modes}

Our experimental data were obtained at $\nu=3/5$. The edge is not chiral at that filling factor and has two upstream neutral modes. On the other hand, the fluctuation-dissipation theorem \cite{WF1,WF2,WF3}, used in the theoretical part of the work, assumes edge chirality. The goal of this Appendix is to clarify under what conditions upstream neutral modes do not undermine the validity of our theoretical analysis. We focus on $t\approx 1$, Fig. 1b). Similar analysis applies to $t\approx 0$. 

The fluctuation-dissipation theorem \cite{WF1,WF2,WF3} was extended to nonchiral systems in Ref. \onlinecite{ups} under the assumption of infinite edges. In practice, edges always terminate in sources and drains. This may lead to the formation of hot spots, where energy  is released due to the equilibration of the incoming charges with the reservoirs. The effect of such hot spots must be eliminated for our results to hold.

The biased charge mode that arrives at the grounded contact G1 in Fig. 4 creates a hot spot h. The spot excites hot upstream neutral modes that may reach ohmic contact D, connected to the amplifier, and generate undesirable additional noise. Our results apply only if such noise is prevented. This can be achieved by introducing a floating ohmic contact F1 as was done in the experiment. The neutral mode dumps its excess energy into F1 and cools to the electron gas temperature.

Another potential problem is the upstream neutral signal, generated at the QPC. The signal may reach the biased contact S and create a hot spot. This may result in undesirable additional noise, emitted from the biased contact. To avoid that a floating ohmic contact F2 is used.

\begin{figure}[b]
	\centering
	\includegraphics[width=3.5in]{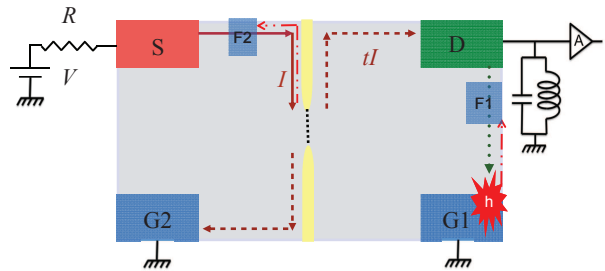}
	\caption{(color online) Hot spot h emits hot upstream neutral modes. Floating ohmic contact F1 cools them down. Floating contact F2 prevents the emergence of a hot spot in the source S.}
	\label{Fig6}
\end{figure}

\end{document}